\documentclass[sigconf]{acmart}
\AtBeginDocument{%
  }

\setcopyright{acmlicensed}
\copyrightyear{2018}
\acmYear{2018}
\acmDOI{XXXXXXX.XXXXXXX}
\acmConference[Conference acronym 'XX]{Make sure to enter the correct
  conference title from your rights confirmation email}{June 03--05,
  2018}{Woodstock, NY}
\acmISBN{978-1-4503-XXXX-X/2018/06}

\usepackage{multirow}
\usepackage{colortbl}  
\usepackage{xcolor}
\usepackage{array}   
\usepackage{marvosym} 
\usepackage{tikz}
\usepackage{pgfplots}
\usepackage{enumitem}
\usepackage{bbm}
\definecolor{linecolor}{gray}{.89} 
\definecolor{linecolor1}{gray}{.93} 
\definecolor{linecolor2}{gray}{.97} 
\usepackage{multirow}
\usepackage{graphicx}
\usepackage[table,xcdraw]{xcolor}
\usepackage{svg}
\usepackage{wrapfig}
\usepackage[most]{tcolorbox}
\usepackage[normalem]{ulem}
\useunder{\uline}{\ul}{}

\begin{document}

\title{Generative Reasoning Recommendation via LLMs}



\author{Minjie Hong}
\email{hongminjie@zju.edu.cn}
\authornote{Both authors contributed equally to this research.}
\affiliation{%
  \institution{Zhejiang University}
  \city{Hangzhou}
  \state{Zhejiang}
  \country{China}
}
\orcid{0009-0000-0368-2527}

\author{Zetong Zhou}
\email{ls2kelvin@gmail.com}
\authornotemark[1]
\affiliation{%
  \institution{Shanghai Jiao Tong University}
  \state{Shanghai}
  \country{China}
}

\author{Zirun Guo}
\email{zrguo.cs@gmail.com}
\affiliation{%
  \institution{Zhejiang University}
  \city{Hangzhou}
  \state{Zhejiang}
  \country{China}
}

\author{Ziang Zhang}
\email{ziangzhang_vsama@qq.com}
\affiliation{%
  \institution{Zhejiang University}
  \city{Hangzhou}
  \state{Zhejiang}
  \country{China}
}

\author{Ruofan Hu}
\email{ruofanhu@zju.edu.cn}
\affiliation{%
  \institution{Zhejiang University}
  \city{Hangzhou}
  \state{Zhejiang}
  \country{China}
}

\author{Weinan Gan}
\email{ganweinan1@huawei.com}
\affiliation{%
  \institution{Huawei Noah's Ark Lab}
  \city{Shenzhen}
  \state{Guangdong}
  \country{China}
}
\orcid{0009-0001-0620-216X}

\author{Jieming Zhu}
\email{jiemingzhu@ieee.org}
\affiliation{%
  \institution{Huawei Noah's Ark Lab}
  \city{Shenzhen}
  \state{Guangdong}
  \country{China}
}
\orcid{0000-0002-5666-8320}

\author{Zhou Zhao}
\email{zhaozhou@zju.edu.cn}
\authornote{Corresponding authors.}
\affiliation{%
  \institution{Zhejiang University}
  \city{Hangzhou}
  \state{Zhejiang}
  \country{China}
}
\orcid{0000-0001-6121-0384}

\renewcommand{\shortauthors}{Hong et al.}

\newcommand{\methodname}{\textsc{GREAM}}
\begin{abstract}
Despite their remarkable reasoning capabilities across diverse domains, large language models (LLMs) face fundamental challenges in natively functioning as \textbf{g}enerative \textbf{r}easoning \textbf{r}ecommendation \textbf{m}odels (\textbf{GRRMs}), where the intrinsic modeling gap between textual semantics and collaborative filtering signals, combined with the sparsity and stochasticity of user feedback, presents significant obstacles. 
This work explores how to build GRRMs by adapting pre-trained LLMs, which achieves a unified \textit{understanding-reasoning-prediction} manner for recommendation tasks. 
Towards this, we propose \textbf{\methodname}, an end-to-end framework that integrates three components: (i) \textbf{Collaborative-Semantic Alignment}, which fuses heterogeneous textual evidence to construct semantically consistent, discrete item indices and auxiliary alignment tasks that ground linguistic representations in interaction semantics; (ii) \textbf{Reasoning Curriculum Activation}, which builds a synthetic dataset with explicit Chain-of-Thought supervision and a curriculum that progresses through behavioral evidence extraction, latent preference modeling, intent inference, recommendation formulation, and denoised sequence rewriting; and (iii) \textbf{Sparse-Regularized Group Policy Optimization (SRPO)}, which stabilizes post-training via \emph{Residual-Sensitive Verifiable Reward} and \emph{Bonus-Calibrated Group Advantage Estimation}, enabling end-to-end optimization under verifiable signals despite sparse successes. Distinct from prior LLM recommenders that trade off efficiency for interpretability, \methodname{} natively supports two complementary inference modes: \emph{Direct Sequence Recommendation} for high-throughput, low-latency deployment, and \emph{Sequential Reasoning Recommendation} that first emits an interpretable reasoning chain for causal transparency. Extensive experiments on three public benchmarks show consistent gains over strong generative and LLM baselines in both direct and reasoning settings. The results demonstrate that \methodname{} achieves a balanced trade-off among efficiency, accuracy, and end-to-end interpretability, providing a practical path for verifiable-RL-driven LLM recommenders. 
The code will be made available at \href{https://github.com/Indolent-Kawhi/GRRM}{https://github.com/Indolent-Kawhi/GRRM}.

\end{abstract}



\received{20 February 2007}
\received[revised]{12 March 2009}
\received[accepted]{5 June 2009}

\maketitle

\section{Introduction}

In web-centric information ecosystems, users face persistent information overload and limited attention. As the backbone of content distribution, recommender systems algorithmically model user behavior to predict and deliver items of potential interest. Traditional two-tower architectures~\cite{wang2021two_tower, hidasi2015gru4rec, li2017neural,kang2018sasrec,johnson2019faiss, guo2020scann} learn low-dimensional embeddings for users and items separately, retrieve candidates via ANN search, and re-rank them with a downstream model. Emerging generative recommendation~\cite{rajput2024tiger,hua2023p5_cid,zhou2025onerec, zhou2025recbasegenerativefoundationmodel} reframes next-item prediction as autoregressive generation (e.g., sequences of discrete item-index tokens), enabling end-to-end learning under a unified language-modeling objective; key challenges include building semantically stable item indices, aligning collaborative semantics, and balancing calibration with efficiency. 

Pushing further, LLM-based recommendation~\cite{hou2024large, cui2022m6, li2023personalized, bao2023tallrec,zheng2024lcrec,hong2025eager, geng2022p5, hua2023p5_cid, liu2024store, zhu2024cllm4rec} imports the understanding–reasoning-prediction capacity of large language models, either by textualizing and instructionalizing behaviors and tasks to exploit linguistic priors, or by semantic indexing that maps items to discrete tokens aligned with the LLM vocabulary—thereby injecting cross-domain knowledge, improving zero-shot generalization, and enhancing interpretability. Nonetheless, a semantic gap between natural-language and collaborative signals, semantic drift from text-only representations, and sparse, verifiable end-to-end training signals continue to limit the attainable performance of LLMs on complex sequential decision-making in recommendation.

Beyond these advances, existing RL paradigms~\cite{tang2025think,shi2025bridging,luo2025llm,zhang2025reinforced,chen2025vragent, gao2025process} for LLM4Rec largely fall into two camps, each with structural drawbacks. (i) Latent-policy or pure-ID optimization: policies are optimized over non-linguistic item identifiers or latent vectors that lack discrete semantic structure. This decouples the policy from the LLM’s pretrained semantics, limits transfer of rich linguistic priors, and yields opaque credit assignment with minimal natural-language supervision. (ii) Text-only reasoning with external retrieval: models produce CoT-style explanations but delegate item selection to an external retriever or a black-box recommender. This breaks end-to-end training, introduces retrieval-induced semantic drift between textual reasoning and final items, and replaces verifiable rewards with exogenous, unverifiable scores—amplifying the unreliability of already sparse signals.

Recently, reinforcement learning with verifiable rewards (RLVR) has delivered notable gains in natural language, mathematical reasoning, and code generation, especially for LLM post-training; methodological advances from PPO\cite{schulman2017proximalpolicyoptimizationalgorithms} through GRPO\cite{shao2024deepseekmathpushinglimitsmathematical}, DAPO~\cite{yu2025dapoopensourcellmreinforcement}, and GSPO~\cite{zheng2025groupsequencepolicyoptimization} have further strengthened complex reasoning. However, directly transplanting RLVR to recommender systems is nontrivial:
\textbf{(1)} Recommendation data are highly structured and the action space often spans millions of items; the semantics learned during LLM pretraining are difficult to align with recommendation objectives and cannot be exhaustively expressed in natural language. 
\textbf{(2)} RLVR hinges on per-sample verifiable rewards, whereas clicks and purchases are counterfactual outcomes conditional on exposure and are confounded by position bias, popularity (Matthew) effects, and feed ordering—thus they cannot be treated as ground-truth labels without careful correction. 
\textbf{(3)} Feedback is extremely sparse, rendering reward/critic learning unstable; many RLVR optimizers presume that each rollout contains both successful and failed samples, a condition rarely satisfied in practice. 
\textbf{(4)} RLVR typically exploits explicit multi-hop discrete reasoning paths (e.g., proofs or execution traces) to support credit assignment, yet the linguistic signals available in recommendation exhibit a substantial semantic gap from such path-structured reasoning, which helps explain the limited benefits observed so far in complex sequential recommendation.

To address the aforementioned challenges, we propose a large language model (LLM)-based generative reasoning recommendation framework that jointly integrates Collaborative-Semantic Alignment, Reasoning Curriculum Activation, and Sparse-Regularized Policy Optimization. This unified framework enables the model to bridge pre-trained linguistic semantics with recommendation-specific collaborative semantics, activate interpretable reasoning capabilities, and perform end-to-end optimization under verifiable rewards.
First, we establish a high-fidelity \textbf{collaborative-semantic alignment} mechanism between the pre-trained semantic space and the recommendation domain. By fusing heterogeneous textual sources—such as item titles, descriptions, and user reviews—we construct semantically consistent and structurally stable discrete item indices. A suite of collaborative alignment tasks is further designed to integrate temporal behavioral patterns and user preferences, ensuring that the model internalizes both linguistic and interaction-level semantics during training.
Second, to endow the model with \textbf{causal and interpretable reasoning ability}, we construct a synthetic reasoning dataset with explicit Chain-of-Thought (CoT) supervision and introduce a curriculum learning strategy that progressively enhances reasoning difficulty. This process guides the model through a structured, five-stage logical chain—behavioral evidence extraction, latent preference modeling, intent inference, recommendation formulation, and denoised sequence rewriting—thus transforming the recommendation task from pattern matching into causal reasoning.
Finally, in the reinforcement learning phase, we propose a \textbf{Sparse-Regularized Group Policy Optimization (SRPO)} algorithm that combines \textbf{Residual-Sensitive Verifiable Reward}, \textbf{Bonus-Calibrated Group Advantage Estimation}. Together, these mechanisms alleviate the instability caused by sparse and stochastic feedback, enhance gradient diversity, and ensure stable policy updates for large-scale recommendation environments.

Unlike previous LLM-based recommendation models that sacrifice interpretability for efficiency, our framework supports two complementary inference modes: \textbf{Direct Sequence Recommendation,} which directly generates the target item index for high-throughput, low-latency inference; \textbf{Sequential Reasoning Recommendation, }where the model first generates an interpretable reasoning chain before producing the final recommendation, enabling causal transparency and verifiable decision-making. This dual-mode design achieves a balanced trade-off between efficiency, accuracy, and interpretability, supporting both large-scale online deployment and research-oriented analyses that require transparent and traceable reasoning.

\begin{itemize}[leftmargin=*]
    \item We introduce an \textbf{LLM-based generative reasoning recommendation framework (\methodname)} that integrates Collaborative-Semantic Alignment and Reasoning Curriculum Activation, and natively supports two inference modes—Direct Sequence and Reasoning-based—to balance efficiency, accuracy, and interpretability.
    \item We develop \textbf{Sparse-Regularized Group Policy Optimization (SRPO)}, coupling Residual-Sensitive Verifiable Reward and Bonus-Calibrated Group Advantage Estimation to stabilize policy learning under sparse feedback. 
    \item Across three public recommendation benchmarks (Beauty, Sports \& Outdoors, Instruments), our method surpasses strong generative and LLM baselines on both direct (Recall/NDCG@K) and reasoning (Pass@K) metrics, demonstrating effectiveness and robustness.
\end{itemize} 
\section{Related Work}
\subsection{LLMs as Recommenders}
Recently, large language models (LLMs) have been adopted for recommendation because their language understanding, generation, and reasoning capabilities enable high-quality textual representations and the incorporation of external knowledge, helping alleviate poor generalization and sparse-interaction regimes ~\cite{liu2024store,lin2023can}.
Two main lines have emerged. (i) Textual/instructional formulations: P5 ~\cite{geng2022p5} unifies multiple recommendation subtasks in an instruction-tuned framework, using a SentencePiece tokenizer to map randomly assigned pseudo-IDs to tokens; M6 ~\cite{cui2022m6} directly uses item text (e.g., names) as identifiers for generative recommendation. While effective for task unification and end-to-end language modeling, these approaches rely solely on text, leading to semantic drift and weak integration of collaborative signals. (ii) Semantic indexing and alignment: LC-Rec ~\cite{zheng2024lcrec} learns discrete IDs via vector quantization and introduces alignment tasks to inject collaborative semantics into LLMs; EAGER-LLM ~\cite{hong2025eager, wang2024eager} proposes a decoder-only generative framework that non-intrusively integrates endogenous/exogenous behavioral and semantic information.
To further improve decision quality, recent work introduces reinforcement learning (RL). LatentR$^3$ ~\cite{zhang2025reinforced} optimizes latent reasoning without CoT data, and Rec-R1 ~\cite{lin2025rec} learns from feedback given by a fixed black-box recommender. Despite gains, two limitations persist: (i) implicit reasoning hampers interpretability and controllability, and (ii) reward signals are sparse/ambiguous or depend on external teachers/retrievers, misaligning the optimization objective with end-to-end generation. In contrast, we introduce an explicit reasoning chain together with a verifiable RL objective on top of LLM-based generative recommendation, enabling truly end-to-end, reasoning-style recommendation.

\subsection{Reinforcement Learning for LLMs}
Reinforcement Learning (RL) has become a cornerstone in advancing large language models (LLMs)~\cite{schulman2017proximalpolicyoptimizationalgorithms, shao2024deepseekmathpushinglimitsmathematical, hong2025apo, hu2025vela, guo2025llm}, often yielding superior generalization compared to purely supervised approaches.The evolution from Trust Region Policy Optimization (TRPO)~\cite{schulman2017trustregionpolicyoptimization} to Proximal Policy Optimization (PPO)~\cite{schulman2017proximalpolicyoptimizationalgorithms}, and later to Group Relative Policy Optimization (GRPO)~\cite{shao2024deepseekmathpushinglimitsmathematical}, marks a pursuit of greater stability, sample efficiency, and scalability for massive model training. Building on these, Decoupled Clip and Dynamic Sampling Policy Optimization (DAPO) introduces decoupled clipping and adaptive sampling to mitigate entropy collapse and improve training stability~\cite{yu2025dapoopensourcellmreinforcement}, while Group Sequence Policy Optimization (GSPO) extends group-based methods by defining importance ratios and optimization at the sequence level, achieving enhanced efficiency and robustness in large-scale RL fine-tuning~\cite{zheng2025groupsequencepolicyoptimization}. These methods have achieved remarkable success in improving reasoning and alignment across diverse domains.
However, when applied to recommendation tasks, post-training RL methods face a significant gap between language pre-training objectives and user-behavior-driven rewards, which are inherently sparse and stochastic. Our method bridges this gap, enabling stable and effective RL-based alignment of LLMs within recommendation scenarios.
\section{Preliminary}
\begin{figure}[t]
  \centering
  \includegraphics[width=0.85\linewidth]{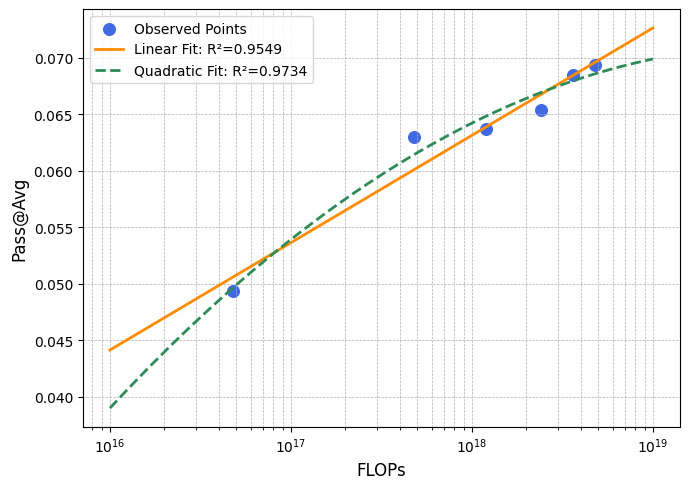}
  \caption{Scaling of reasoning performance vs. Reasoning Activation training compute. 
    \textbf{Pass@Avg increases almost linearly with compute and shows no sign of saturation}, indicating that our method has not yet reached its performance ceiling.
    }
  \label{fig:scaling_law}
\end{figure}

\subsection{Generative Recommendation. }
\subsubsection{\textbf{\textit{Problem Formulation. }}} Sequential recommendation is a crucial metric in LLM-based recommender systems. We transform the traditional two-tower model, which computes similarity followed by reordering, into a generative recommendation paradigm. In this framework, each item  $\mathbf{x}$  is represented by a set of tokens ${\bf Y}=[{\bf y}_1, {\bf y}_2, \cdots, {\bf y}_k]\in \mathcal{Y}$. Given an input sequence  $\mathbf X$, which includes instructions and the interaction history, the sequence of the target item  $\mathbf{Y}$  is generated directly in an autoregressive manner. The probability can be calculated by:
\begin{equation}
    p({\bf Y}|{\bf X}) = {\prod}_{i=1}^{k} p({\bf y}_i|\bf{X},{\bf y}_1,{\bf y}_2,\dots,{\bf y}_{i-1})
\end{equation}
This end-to-end formulation replaces retrieval–re-ranking and directly optimizes representation, generation, and decision.

\subsubsection{\textbf{\textit{Knowledge-rich Item Indices. }}}

LLM as an item semantic extractor. For each item $i$, feed its text $x_i$ to an LLM (parameters $\theta$) to obtain the final-layer hidden states
\begin{equation}
\mathbf{H}_i^{(L)} \;=\; [\,\mathbf{h}_{i,1},\dots,\mathbf{h}_{i,L_i}\,] \in \mathbb{R}^{L_i \times d},
\end{equation}
and take the last hidden state as the item embedding
\begin{equation}
\mathbf{e}_i \;:=\; \mathbf{h}_{i,L_i} \in \mathbb{R}^{d}.
\end{equation}
\textbf{RQ-KMeans residual quantization. }Let $H$ be the number of levels and $N_l$ the level-$l$ codebook size. Initialize
\begin{equation}
\mathbf{R}_i^{(1)} \;=\; \mathbf{e}_i.
\end{equation}
For each level $l=1,\dots,H$, run K-means on $\{\mathbf{R}_i^{(l)}\}_{i\in\mathcal{I}}$ to obtain
\begin{equation}
\mathcal{C}^{(l)} \;=\; \mathrm{K\text{-}means}\!\big(\{\mathbf{R}_i^{(l)}\},\,N_l\big)
=\big\{\mathbf{c}^{(l)}_{k}\,\big|\,k=1,\dots,N_l\big\}.
\end{equation}
assign the nearest centroid index
\begin{equation}
s_i^{(l)} \;=\; \arg\min_{k\in\{1,\dots,N_l\}} \big\|\,\mathbf{R}_i^{(l)} - \mathbf{c}^{(l)}_{k}\,\big\|,
\end{equation}
and update the residual
\begin{equation}
\mathbf{R}_i^{(l+1)} \;=\; \mathbf{R}_i^{(l)} - \mathbf{c}^{l}_{\,s_i^{(l)}}.
\end{equation}
The hierarchical discrete index is
\begin{equation}
\mathbf{s}_i \;=\; \big(s_i^{(1)},\dots,s_i^{(H)}\big), \qquad s_i^{(l)} \in \{1,\dots,N_l\}.
\end{equation}
We map $\mathbf{s}_i$ to LLM-vocabulary tokens (e.g., $\langle a\_053\rangle\,\langle b\_023\rangle\,\cdots$). The residual hierarchy induces prefix sharing and thus semantic proximity: longer common prefixes imply stronger coarse-grained agreement, benefiting interpretability and coarse-to-fine training.

\subsection{Reinforcement Learning with Verifiable Reward.}
Reinforcement learning aims to maximize the following objective:
\begin{equation}\label{eq1}
    \mathcal{J}(\theta)=\mathbb{E}_{y\thicksim\pi_\theta}[r(y)-\beta \mathbb{D}_{\mathrm{KL}}(\pi_\theta||\pi_{\theta_{\text{ref}}})]
\end{equation}
where $y$ denotes the generated response, $r(\cdot)$ is the reward function, $\mathbb{D}_{\mathrm{KL}}$ represents the KL divergence, $\beta$ controls the penalty trade-off, and $\pi_\theta$ and $\pi_{\theta_\text{ref}}$ denote the policy and reference models, respectively. In GRPO~\cite{shao2024deepseekmathpushinglimitsmathematical}, the advantage is estimated in a group-relative manner. Concretely, for a question-answer pair $(q,a)$, the policy $\pi_\theta$ samples a group of $G$ distinct responses $\{o_i\}_{i=1}^G$ and then the advantage is computed as:
\begin{equation}
    \hat{A}_{i}=\frac{r_i-\mathrm{mean}(\{r_i\}_{i=1}^G)}{\mathrm{std}(\{r_i\}_{i=1}^G)}
\end{equation}
where $r_i$ is the rule-based verifiable reward~\citep{shao2024deepseekmathpushinglimitsmathematical}. Following PPO~\cite{schulman2017proximalpolicyoptimizationalgorithms}, GRPO adopts a clipped objective.
Let the importance sampling ratio be defined as:
\begin{equation}\label{eq:isratio}
r_{i,t}(\theta)=\frac{\pi_\theta(o_{i,t}\mid q,o_{i,<t})}{\pi_{\theta_{\mathrm{old}}}(o_{i,t}\mid q,o_{i,<t})},
\end{equation}
then the GRPO objective in Eq.~\ref{eq1} can be reformulated as:

\begin{equation}\label{eqgrpo}
\begin{aligned}
&\mathcal{J}_{\mathrm{GRPO}}(\theta)
= \mathbb{E}_{(q,a)\sim\mathcal{D},\,\{o_i\}_{i=1}^G\sim\pi_{\theta_{\mathrm{old}}}(\cdot|q)} \\
&\qquad\qquad 
\Biggl[
\frac{1}{G}\sum_{i=1}^{G}\frac{1}{|o_{i}|}\sum_{t=1}^{|o_{i}|}
\Bigl(
\min\bigl(r_{i,t}(\theta)\hat{A}_{i,t},\, \\
&\qquad\qquad 
\mathrm{clip}\bigl(r_{i,t}(\theta),\, 1-\varepsilon,
1+\varepsilon\bigr)\hat{A}_{i,t}\bigr) \\
&\qquad\qquad 
-\beta\, \mathbb{D}_{\mathrm{KL}}(\pi_{\theta}\,\|\,\pi_{\mathrm{ref}})
\Bigr)
\Biggr]
\end{aligned}
\end{equation}

%
\section{Collaborative-Semantic Alignment and Reasoning Activation}
\label{sec:semantic_alignment}

\begin{figure*}[t]
  \centering
  \includegraphics[width=0.85\linewidth]{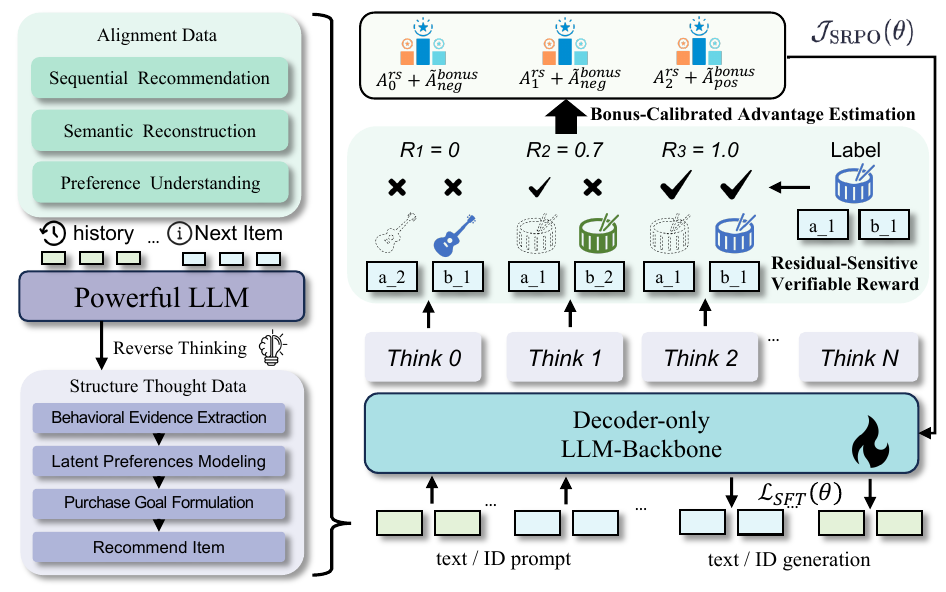}
  \caption{An overall framework of our \textbf{\methodname}. \textbf{(Left)}: We construct alignment data and synthesize structured reasoning data via a reverse-reasoning process to perform supervised fine-tuning on LLM. \textbf{(Right)}: We employ SRPO for further reinforcement learning training , which enables LLM to learn more effectively from sparse reward signals through our proposed Residual-sensitive Reward and Bonus-Calibrated Advantage.}
  \label{fig:overall_framework}
\end{figure*}

The successful application of LLMs to recommendation requires bridging the significant gap between their general-purpose, pre-trained semantic space and the specialized, collaborative semantics inherent in user-item interaction data. A model trained solely on textual data risks semantic drift, while one relying only on collaborative signals fails to harness the rich knowledge and reasoning capabilities of LLMs. To address this, we introduce a preparatory stage that constructs two distinct yet synergistic types of data, designed to (i) establish a high-fidelity, semantically-aligned item representation space and (ii) activate the model's latent reasoning abilities for complex sequential decision-making.

\subsection{High-Fidelity Indexing and Collaborative Alignment Data}
\label{subsec:alignment_data}

A core challenge in generative recommendation is creating discrete item indices that are both semantically meaningful and computationally efficient. To this end, we first enhance the quality of our item representations. Recognizing that item titles are often concise but lack detail, while user reviews provide rich, user-centric perspectives, we synthesize these sources. For each item, we aggregate its title, official description, and a curated set of high-quality user reviews. This composite text is then processed by a powerful, state-of-the-art generative model (i.e., GPT-5) instructed to rewrite a comprehensive, feature-rich description that encapsulates both objective attributes and subjective user perceptions. By encoding the concatenation of the original item title and this newly synthesized description through our LLM, we produce embeddings that are substantially more informative. Applying RQ-KMeans~\cite{luo2024qarm} to these enriched embeddings significantly reduces semantic ambiguity, leading to an approximate \textbf{50\% reduction in index collision rates} compared to using titles alone.

With these high-fidelity indices, we then construct the alignment dataset, $\mathcal{D}_{\text{align}}$, by formulating a diverse suite of tasks designed to instill collaborative knowledge. These tasks fall into three main categories:

\begin{itemize}[leftmargin=*]
    \item \textbf{Sequential Recommendation Tasks:} These tasks frame next-item prediction as a language modeling problem (e.g., given a user's history, predict the index of the next item). This teaches the model the temporal dynamics and sequential patterns of user behavior.

    \item \textbf{Semantic Reconstruction Tasks:} These tasks create a strong link between textual semantics and our discrete indices. Examples include predicting an item's tokenized index from its textual description, and vice versa. This ensures the model's internal representations are well-grounded in both collaborative and semantic spaces.

    \item \textbf{User Preference-related Tasks:} These tasks focus on modeling a user's general, long-term interests. For instance, the model might be asked to generate a textual summary of a user's preferences based on their interaction history.
\end{itemize}

This multiscale alignment dataset ensures that the model understands not just \textit{what} an item is, but also \textit{how} it fits within the broader ecosystem of user behaviors and collaborative patterns.

\subsection{Reasoning Activation via Synthetic Chain-of-Thought Data}
\label{subsec:reasoning_data}

To elevate the model from a pattern-matching predictor to a veritable reasoner, we construct a high-quality, synthetic dataset, $\mathcal{D}_{\text{reason}}$, designed to explicitly teach Chain-of-Thought (CoT) reasoning for recommendation. A critical insight guiding our approach is that sequential recommendation is not a deterministic task with a single ``correct'' answer; the observed next item is merely one of many plausible outcomes. Therefore, the primary goal of our data synthesis is not to perfectly replicate historical sequences but to forge a \textbf{plausible and interpretable causal link} between a user's interaction history and a subsequent, recommended item.

We achieve this through a reverse-reasoning process, employing a powerful LLM to generate structured reasoning paths that culminate in a logical recommendation. Our framework follows the reasoning template proposed in DeepSeek-R1, where system prompts enforce output-format constraints (e.g., <think>...</think> and <answer>...</answer>), ensuring structural consistency between reasoning logic and the generated recommendation results. Specifically, we redefine the Reason component of the sequential reasoning recommendation task as a standardized logical chain composed of five stages:

\begin{itemize}[leftmargin=*]
    \item \textbf{Extraction of Behavioral Evidence:} The model first analyzes the user's historical interaction sequence to identify salient patterns, such as evolving interests, functional preferences (e.g., preference for a certain brand), and underlying motivations.

    \item \textbf{Modeling of Latent Preferences:} Based on the extracted evidence, the model infers a coherent user persona, including long-term characteristics, brand affinities, and value orientations, forming a causally interpretable profile.

    \item \textbf{Inference of User Intent and Goal Formulation:} The model synthesizes the user persona with recent behaviors to deduce the user's current latent need and formulate a specific goal for the next interaction (e.g., ``find a complementary product'' or ``upgrade an existing item'').

    \item \textbf{Formulation of Recommendation and Justification:} The model generates a concrete item recommendation and provides an explicit justification that connects the recommendation directly back to the inferred intent and historical evidence, ensuring transparency.

    \item \textbf{Denoised Sequence Rewriting:} As a final step, the model learns to autonomously refine the input history by identifying and filtering out noisy or irrelevant interactions. This enhances policy robustness by enabling the model to focus on the most signal-rich elements of a user's sequence.
\end{itemize}

This structured CoT data effectively transforms the recommendation task from a black-box prediction into a transparent, step-by-step reasoning problem, activating the LLM's inherent capacity for complex logical deduction. Specific case studies that show the concrete responses of \methodname can be found in Appendix \ref{casestudy}.

\subsection{Hybrid SFT with Curriculum Learning}
\label{subsec:hybrid_sft}

We discovered that the alignment data ($\mathcal{D}_{\text{align}}$) and the reasoning data ($\mathcal{D}_{\text{reason}}$) are highly synergistic. The former grounds the model in the factual, collaborative structure of the item space, while the latter teaches it how to navigate this space through logical inference. Consequently, we train the model using Supervised Fine-Tuning (SFT) on a hybrid mixture of both datasets, $\mathcal{D}_{\text{SFT}} = \mathcal{D}_{\text{align}} \cup \mathcal{D}_{\text{reason}}$. The training objective is to minimize the standard auto-regressive negative log-likelihood loss:
\[
\mathcal{L}_{\text{SFT}}(\theta) = -\sum_{(X, Y) \in \mathcal{D}_{\text{SFT}}} \sum_{t=1}^{|Y|} \log p_{\theta}(y_t \mid y_{<t}, X)
\]
where $(X, Y)$ is an input-output pair from the combined dataset, and $\theta$ represents the model parameters. The resulting model is versatile, capable of performing both direct, low-latency item ID generation and more deliberative, reasoning-backed recommendation where the model first produces an interpretable CoT before outputting the final item ID.

To maximize the benefits of this hybrid approach, we employ a curriculum learning strategy over three epochs. A naive uniform mixing can be suboptimal, as the model may struggle with complex reasoning without a solid semantic foundation. Our curriculum allows the model to first establish a strong grasp of collaborative semantics before progressively focusing on reasoning tasks. During the first two epochs, for each alignment batch $b_{\text{align}, i} \in B_{\text{align}}$ processed, we insert a reasoning batch from $B_{\text{reason}}$ with a probability $p_{\text{insert}}(i)$ that increases with the training step $i$:
\[
p_{\text{insert}}(i) = \min\left(1, \gamma \cdot \frac{i}{N_{\text{align}}} \cdot \frac{N_{\text{reason}}}{N_{\text{align}}}\right),
\]
where $N_{\text{align}} = |B_{\text{align}}|$ and $N_{\text{reason}} = |B_{\text{reason}}|$ are the total number of batches for each data type, and $\gamma$ is a hyperparameter scaling the curriculum pace (empirically set to 1.5). Any remaining reasoning batches are appended at the end of the epoch to ensure all data is utilized. In the final epoch, we switch to a uniform sampling distribution, mixing the two data types evenly to fine-tune both capabilities concurrently. 

\section{Sparse-Regularized Group Policy Optimization (SRPO)}

\subsection{Residual-sensitive Verifiable Reward Shaping}
In our residual discretization indexing scheme, higher-level indices encode coarser-grained and more discriminative semantics, where prefix consistency directly determines both the semantic correctness of the neighborhood and the contraction speed of the candidate space.  
Traditional binary rewards (i.e., ``completely correct or completely wrong'') in generative recommendation often suffer from reward sparsity and hinder accurate credit assignment.  
To address this issue, we propose a \textbf{residual-sensitive verifiable reward shaping} mechanism that leverages the longest common prefix length between the generated and target indices as a computable and verifiable fine-grained signal, ensuring goal consistency while substantially alleviating reward sparsity.

Let the target item index be $y = (c_1, \dots, c_H)$,
and the model-generated index be $\hat{y} = (\hat{c}_1, \dots, \hat{c}_H)$,
where $H$ denotes the number of hierarchical levels (e.g., $H=4$).  
We define the longest common prefix length as:
\begin{equation}
\ell = \max \{ k \in \{0, \dots, H\} \mid \hat{c}_i = c_i,\ \forall i \le k \}.
\end{equation}
Accordingly, the \textbf{residual-sensitive verifiable reward} is formulated as:
\begin{equation}
r^{\text{rs}} = \left( \frac{\ell}{H} \right)^{\beta},
\end{equation}
where $\beta \in (0, 1]$ controls the concavity of the reward curve.  
We set $\beta = \tfrac{1}{2}$ by default.

This reward design provides four key advantages.  
First, it introduces \textbf{densification and stability}: whenever $\ell > 0$, the model receives a non-zero reward, effectively mitigating sparsity and reducing variance compared to binary all-or-nothing signals.  
Second, it highlights \textbf{prefix importance}: since $\beta \in (0,1)$ makes $r_{\text{rs}}$ a concave function, each incremental prefix match ($\Delta \ell = 1$) yields a larger marginal gain at earlier levels, aligning with the hierarchical property that upper-level indices encode coarser and more critical semantics, thereby assigning higher implicit weights to earlier indices.  
Third, it facilitates \textbf{curriculum optimization}, as the training process naturally follows a coarse-to-fine trajectory—first capturing semantic categories at higher levels and progressively refining to instance-level representations.  
Finally, it ensures \textbf{verifiability}: the computation of $\ell$ is deterministic and unambiguous, satisfying the verifiable reward criterion of RLVR without requiring additional annotations or external retrieval modules.

\subsection{Bonus-Calibrated Group Advantage Estimation}

In the early stages of generative recommendation, exactly correct generations are typically rare. While the residual-sensitive dense reward $r^{\mathrm{rs}}$ supplies stable gradients, it may not sufficiently drive these rare success events. To address this, beyond GRPO's group-relative advantage, we introduce a group success--oriented bonus term that explicitly optimizes the objective ``at least one generation is exactly correct.'' Inspired by \textsc{Pass@k}~\cite{chen2025pass}, we measure this target via the group success probability under without-replacement sampling. However, unlike using \textsc{Pass@k} alone, we note a key limitation: when the sample size exceeds the number of incorrect generations, the success probability $\rho$ saturates toward $1$ and its variance collapses toward $0$, causing all advantages to approach zero and hindering further optimization. To avoid this degeneration, we add this bonus term directly to the residual dense advantage, yielding complementary signals that are dense--stable (from $A^{\mathrm{rs}}_i$) and rare--exact (from the bonus).

Let $G$ be the rollout size, and let $v_i \in \{0,1\}$ indicate whether the $i$-th generation is exactly correct ($v_i=1$) or not ($v_i=0$). Denote $c=\sum_{i=1}^{G} v_i$ as the count of exactly correct generations. For without-replacement sampling of size $k$ with $1 \le k \le G$, the group success probability that at least one of the $k$ samples is exactly correct is
\begin{equation}
\label{eq:rho}
\rho(c;G,k) \;=\; 1 \;-\; \frac{\binom{G-c}{k}}{\binom{G}{k}} \, .
\end{equation}
where $\binom{n}{m}$ denotes the binomial coefficient. We also define the group-level Bernoulli variance
\begin{equation}
\label{eq:sigma}
\sigma \;=\; \sqrt{\rho\,(1-\rho)} \, ,
\end{equation}
\textbf{Dense–stable advantage.} The group-normalized residual advantage is
\begin{equation}
\label{eq:rs-adv}
A^{rs}_i \;=\; \frac{\, r^{\mathrm{rs}}_i \;-\; \mathrm{mean}\!\big(\{ r^{\mathrm{rs}}_i \}_{i=1}^{G} \big) \,}{\, \mathrm{std}\!\big(\{ r^{\mathrm{rs}}_i \}_{i=1}^{G} \big) \,} \, .
\end{equation}
\textbf{Rare–exact bonus advantage.} For \emph{exactly correct} samples ($v_i=1$), we assign the positive bonus
\begin{equation}
\label{eq:bonus-pos}
\tilde{A}^{\mathrm{bonus}}_{\mathrm{pos}} \;=\; \frac{\, 1 - \rho \,}{\, \sigma \,} \, ,
\end{equation}
which grows with the \emph{gap-to-perfection} $(1-\rho)$. For \emph{incorrect} samples ($v_i=0$), we consider their counterfactual marginal contribution to group success if flipped to correct. Let
\begin{equation}
\label{eq:delta}
\delta \;=\; \frac{\binom{G-c-1}{\,k-1\,}}{\binom{G-1}{\,k-1\,}} \, ,
\end{equation}
the probability that this sample is selected and the remaining $k-1$ draws are all incorrect. We then define the negative bonus
\begin{equation}
\label{eq:bonus-neg}
\tilde{A}^{\mathrm{bonus}}_{\mathrm{neg}} \;=\; \frac{\, (1-\rho) - \delta \,}{\, \sigma \,} \, .
\end{equation}
Together, \eqref{eq:bonus-pos}--\eqref{eq:bonus-neg} yield a group-level counterfactual credit assignment: positives share the group's gap-to-perfection, while negatives receive signals proportional to their potential flip-to-success contributions.

To prevent rare–exact bonus advantage saturation ($\rho \to 1$, $\sigma \to 0$) from collapsing all advantages while preserving the dense--stable properties of $A^{rs}_i$, we form the final advantage by direct addition:
\begin{equation}
A_i^{\mathrm{final}}
\;=\;
A^{rs}_i
\;+\;
\mathbbm{1}[v_i=1]\;\tilde{A}^{\mathrm{bonus}}_{\mathrm{pos}}
\;+\;
\mathbbm{1}[v_i=0]\;\tilde{A}^{\mathrm{bonus}}_{\mathrm{neg}} \, ,
\end{equation}
where $\mathbbm{1}[\cdot]$ is the indicator function. This ensures that early training benefits from the stability and prefix alignment of $A^{rs}_i$, while the bonus $\tilde{A}^{\mathrm{bonus}}$ strongly drives exactly correct events later on; even when $k > G-c$ yields $\rho \approx 1$ and $\sigma \to 0$, overall learning does not collapse because $A^{rs}_i$ remains active.

\subsection{Final Objective.}

Recommendation rewards are typically sparse. While DAPO’s dynamic sampling mitigates the gradient-decreasing pathology whereby identical group rewards yield zero group-normalized advantages and thus vanishing policy updates, its criterion is tailored to 0/1 accuracies. In our setting, per-rollout rewards are continuous. We therefore adopt a variance-guided dynamic sampling strategy: reject groups with sampling variance equal to zero. Then the objective can be reformulated as:

\begin{equation}
\begin{aligned}
&\mathcal{J}_{\mathrm{SRPO}}(\theta)
= \mathbb{E}_{(q,a)\sim\mathcal{D},\,\{o_i\}_{i=1}^G\sim\pi_{\theta_{\mathrm{old}}}(\cdot|q)} \\
&\qquad\qquad 
\Biggl[
\frac{1}{G}\sum_{i=1}^{G}\frac{1}{|o_{i}|}\sum_{t=1}^{|o_{i}|}
\Bigl(
\min\bigl(r_{i,t}(\theta){\color{red}A^{\mathrm{final}}_{i,t}},\, \\
&\qquad\qquad 
\mathrm{clip}\bigl(r_{i,t}(\theta),\, 1-\varepsilon,
1+\varepsilon\bigr){\color{red}A^{\mathrm{final}}_{i,t}}\bigr) \\
&\qquad\qquad 
-\beta\, \mathbb{D}_{\mathrm{KL}}(\pi_{\theta}\,\|\,\pi_{\mathrm{ref}})
\Bigr)
\Biggr] \\
&\qquad \text{s.t.      } {\color{red}\mathrm{std}\!\big(\{ r^{\mathrm{rs}}_i \}_{i=1}^{G} \big) \neq 0}.
\end{aligned}
\end{equation}
\section{EMPIRICAL STUDY}


\begin{table}[tp]
\centering
\caption{Statistics of the Datasets.}
\label{tab:dataset}
\resizebox{\linewidth}{!}{
\begin{tabular}{ccccc}
\toprule
Dataset & \#Users & \#Items & \#Interactions & \#Sparsity \\ \midrule
Beauty &22,363  & 12,101  & 198,360 & 0.00073\\
Sports and Outdoors &35,598  & 18,357  &296,175 & 0.00045\\
Instruments &24,733  & 9,923  & 206,153 & 0.00083\\
\bottomrule
\end{tabular}
}
\end{table}
\begin{table*}[]
\centering
\resizebox{\textwidth}{!}
{%
\begin{tabular}{
>{\columncolor[HTML]{FFFFFF}}c 
>{\columncolor[HTML]{FFFFFF}}l 
>{\columncolor[HTML]{FFFFFF}}c 
>{\columncolor[HTML]{FFFFFF}}c 
>{\columncolor[HTML]{FFFFFF}}c 
>{\columncolor[HTML]{FFFFFF}}c 
>{\columncolor[HTML]{FFFFFF}}c 
>{\columncolor[HTML]{FFFFFF}}c 
>{\columncolor[HTML]{FFFFFF}}c 
>{\columncolor[HTML]{FFFFFF}}c 
>{\columncolor[HTML]{FFFFFF}}c 
>{\columncolor[HTML]{FFFFFF}}c |
>{\columncolor[HTML]{EFEFEF}}c 
>{\columncolor[HTML]{EFEFEF}}c |l
>{\columncolor[HTML]{EFEFEF}}c 
>{\columncolor[HTML]{EFEFEF}}c }
\hline
\cellcolor[HTML]{FFFFFF} & \multicolumn{1}{c}{\cellcolor[HTML]{FFFFFF}} & \multicolumn{3}{c}{\cellcolor[HTML]{FFFFFF}Traditional} & \multicolumn{3}{c}{\cellcolor[HTML]{FFFFFF}Transformer-based} & \multicolumn{2}{c}{\cellcolor[HTML]{FFFFFF}Generative} & \multicolumn{2}{c|}{\cellcolor[HTML]{FFFFFF}LLM-based} & \multicolumn{2}{c|}{\cellcolor[HTML]{FFFFFF}\methodname} & \multicolumn{1}{c}{\cellcolor[HTML]{FFFFFF}} & \multicolumn{2}{c}{\cellcolor[HTML]{FFFFFF}\methodname} \\ \cline{3-14} \cline{16-17} 
\multirow{-2}{*}{\cellcolor[HTML]{FFFFFF}Dataset} & \multicolumn{1}{c}{\multirow{-2}{*}{\cellcolor[HTML]{FFFFFF}\begin{tabular}[c]{@{}c@{}}Direct\\ Metric\end{tabular}}} & GRU4REC & Caser & HGN & Bert4Rec & S\textsuperscript{3}-Rec & FDSA & P5-CID & TIGER & LC-Rec\textsuperscript{*} & EAGER-LLM\textsuperscript{*} & Align & RL & \multicolumn{1}{c}{\multirow{-2}{*}{\cellcolor[HTML]{FFFFFF}\begin{tabular}[c]{@{}c@{}}Reason \\ Metric\end{tabular}}} & Align & RL \\ \hline
\cellcolor[HTML]{FFFFFF} & Recall@1 & 0.0571 & 0.0149 & 0.0523 & 0.0435 & 0.0367 & 0.0520 & 0.0587 & 0.0608 & 0.0656 & 0.0680 & \textbf{0.0711} & {\ul 0.0689} & Pass@1 & 0.0495 & {\color[HTML]{333333} \textbf{0.0650}} \\
\cellcolor[HTML]{FFFFFF} & Recall@5 & 0.0821 & 0.0543 & 0.0813 & 0.0671 & 0.0863 & 0.0834 & 0.0827 & 0.0863 & 0.0920 & {\ul 0.0963} & \textbf{0.1026} & 0.0957 & Pass@5 & 0.0705 & \textbf{0.0765} \\
\cellcolor[HTML]{FFFFFF} & Recall@10 & 0.1031 & 0.0710 & 0.1048 & 0.0822 & 0.1136 & 0.1046 & 0.1016 & 0.1064 & 0.1115 & {\ul 0.1171} & \textbf{0.1207} & 0.1139 & Pass@10 & 0.0829 & \textbf{0.0845} \\
\cellcolor[HTML]{FFFFFF} & NDCG@5 & 0.0698 & 0.0355 & 0.0668 & 0.0560 & 0.0626 & 0.0681 & 0.0708 & 0.0738 & 0.0790 & 0.0823 & \textbf{0.0872} & {\ul 0.0825} & - & - & \textbf{-} \\
\multirow{-5}{*}{\cellcolor[HTML]{FFFFFF}Instruments} & NDCG@10 & 0.0765 & 0.0409 & 0.0744 & 0.0608 & 0.0714 & 0.0750 & 0.0768 & 0.0803 & 0.0853 & {\ul 0.0890} & \textbf{0.0931} & 0.0884 & - & - & \textbf{-} \\ \hline
\cellcolor[HTML]{FFFFFF} & Recall@1 & - & - & - & - & - & - & - & - & 0.0107 & 0.0105 & \textbf{0.0120} & {\ul 0.0110} & Pass@1 & 0.0043 & \textbf{0.0074} \\
\cellcolor[HTML]{FFFFFF} & Recall@5 & 0.0129 & 0.0116 & 0.0189 & 0.0115 & 0.0251 & 0.0182 & 0.0313 & 0.0264 & 0.0302 & 0.0349 & \textbf{0.0372} & {\ul 0.0355} & Pass@5 & 0.0163 & \textbf{0.0201} \\
\cellcolor[HTML]{FFFFFF} & Recall@10 & 0.0204 & 0.0194 & 0.0313 & 0.0191 & 0.0385 & 0.0288 & 0.0431 & 0.0400 & 0.0465 & {\ul 0.0555} & \textbf{0.0556} & 0.0523 & Pass@10 & 0.0275 & \textbf{0.0300} \\
\cellcolor[HTML]{FFFFFF} & NDCG@5 & 0.0086 & 0.0072 & 0.0120 & 0.0075 & 0.0161 & 0.0122 & 0.0224 & 0.0181 & 0.0193 & 0.0227 & \textbf{0.0247} & {\ul 0.0234} & - & - & \textbf{-} \\
\multirow{-5}{*}{\cellcolor[HTML]{FFFFFF}Sports} & NDCG@10 & 0.0110 & 0.0097 & 0.0159 & 0.0099 & 0.0204 & 0.0156 & 0.0262 & 0.0225 & 0.0247 & {\ul 0.0293} & \textbf{0.0307} & 0.0289 & - & - & \textbf{-} \\ \hline
\cellcolor[HTML]{FFFFFF} & Recall@1 & - & - & - & - & - & - & - & - & 0.0143 & 0.0171 & \textbf{0.0190} & {\ul 0.0172} & Pass@1 & 0.0079 & \textbf{0.0137} \\
\cellcolor[HTML]{FFFFFF} & Recall@5 & 0.0164 & 0.0205 & 0.0325 & 0.0203 & 0.0387 & 0.0267 & 0.0400 & 0.0454 & 0.0494 & 0.0534 & \textbf{0.0567} & {\ul 0.0551} & Pass@5 & 0.0270 & \textbf{0.0296} \\
\cellcolor[HTML]{FFFFFF} & Recall@10 & 0.0283 & 0.0347 & 0.0512 & 0.0347 & 0.0647 & 0.0407 & 0.0590 & 0.0648 & 0.0740 & {\ul 0.0787} & \textbf{0.0814} & 0.0771 & Pass@10 & \textbf{0.0446} & 0.0403 \\
\cellcolor[HTML]{FFFFFF} & NDCG@5 & 0.0099 & 0.0131 & 0.0206 & 0.0124 & 0.0244 & 0.0163 & 0.0274 & 0.0321 & 0.0330 & 0.0363 & \textbf{0.0383} & {\ul 0.0365} & - & - & \textbf{-} \\
\multirow{-5}{*}{\cellcolor[HTML]{FFFFFF}Beauty} & NDCG@10 & 0.0137 & 0.0176 & 0.0266 & 0.0170 & 0.0327 & 0.0208 & 0.0335 & 0.0384 & 0.0417 & {\ul 0.0451} & \textbf{0.0463} & 0.0436 & - & - & \textbf{-} \\ \hline
\end{tabular}%
}
\caption{Performance comparison of various recommendation methods. The best and second-best results are marked in bold and \underline{underline}, respectively. Direct denotes the sequence direct recommendation task, whereas Reason corresponds to the sequential reasoning recommendation task. For fairness, both LC-Rec\textsuperscript{*} and EAGER-LLM\textsuperscript{*} were re-trained under identical configurations using the Qwen3-4B-Instruct backbone, ensuring consistent experimental conditions and reproducibility. }
\label{tab:main_result}
\end{table*}
\subsection{Experimental Setting}
\subsubsection{\textbf{\textit{Dataset.}}}
To comprehensively evaluate our approach, we conducted experiments on three real-world public datasets of Amazon product reviews~\cite{mcauley2015amazon, he2016amazon_2}, which are widely recognized benchmarks in sequence recommendation research. Specifically, we used data from three subcategories—Beauty, Sports and Outdoors, and Musical Instruments. Consistent with prior work~\cite{hou2022towards, rendle2010factorizing, zhang2019feature}, we applied the 5-core filtering protocol, excluding users and items with fewer than five interactions to retain meaningful behavioral sequences. The detailed dataset statistics are provided in Table~\ref{tab:dataset}. In addition, we leverage multiple open- and closed-source large language models (e.g., Qwen3, GPT-5, and Llama-3.1) to generate large-scale textual data based on these datasets for alignment purposes. The structure of the generated data can be referred to in the case study provided in the Appendix \ref{casestudy}.

\subsubsection{\textbf{\textit{Evaluation Metrics.}}}

We follow standard evaluation protocols~\cite{kang2018sasrec} and adopt the leave-one-out strategy for assessment. For direct sequence recommendation tasks, we employ two widely recognized metrics—Recall and Normalized Discounted Cumulative Gain (NDCG)—and report results at cutoffs of k = 5 and k = 10. For generative recommendation methods utilizing beam search, the beam size is consistently fixed at 20. In the case of sequential recommendation tasks, we evaluate performance using Pass@k accuracy at k = 1, 5, and 10, with the generation temperature set to 0.7 and the top-p value to 0.9. 

\subsubsection{\textbf{\textit{Implementation Details.}}}
All experiments are conducted on a cluster equipped with 24 NVIDIA A100 GPUs. We adopt Qwen3-4B-Instruct-2507 as the backbone model. \textbf{The item indexing scheme} is designed with a fixed length of 5, where the fifth position serves as a conflict-resolution slot, and each hierarchical layer is assigned a structured identifier capable of representing 256 distinct values. During \textbf{Collaborative-Semantic Alignment and
Reasoning Activation phase}, we employ multiple large language models—Qwen3-30B, GPT-5, Llama-3.1, and Gemini—to synthesize domain-diverse textual data and mitigate overfitting. In the \textbf{alignment phase}, the model is fully fine-tuned using the LLaMA-Factory framework, with a learning rate of 5.0e-5 for 3 epochs. For \textbf{reinforcement learning optimization}, we adopt the Verl~\cite{sheng2024hybridflow} framework with the following hyperparameters: learning rate = 4e-6, weight decay = 0.01, training batch size = 256, mini-batch size = 128, and KL-loss coefficient = 0.001. The rollout configuration sets temperature = 0.5, top-p = 1.0, and samples 10 responses per query. We utilize SGLang as both the RL rollout backend and the inference acceleration framework for \textbf{evaluation}.

\subsection{Performance Comparison}
\begin{table}[]
\centering
\resizebox{\columnwidth}{!}{%
\begin{tabular}{lcc}
\hline
\multicolumn{1}{c|}{Variants} & \begin{tabular}[c]{@{}c@{}}Avg Direct\\ Metric\end{tabular} & \begin{tabular}[c]{@{}c@{}}Avg Reason\\ Metric\end{tabular} \\ \hline
\multicolumn{3}{c}{\cellcolor[HTML]{EFEFEF}Align} \\
+ Collaborative-Semantic Alignment & 0.0865 & - \\
+ Reasoning Curriculum Activation & \textbf{0.0949} & 0.0656 \\
\multicolumn{3}{c}{\cellcolor[HTML]{C0C0C0}RL} \\
GRPO & 0.0921 & 0.0741 \\
+ Residual-sensitive Verifiable Reward & 0.0891 & 0.0724 \\
+ Bonus-Calibrated Advantage Estimation & 0.0899 & \textbf{0.0753} \\ \hline
\end{tabular}%
}
\caption{Ablation of Collaborative-Semantic Alignment, Reasoning Activation and Sparse-Regularized Group Policy Optimization(SRPO) Components evaluated by Avg Direct (mean Recall/NDCG@1/5/10) and Avg Reason (mean Pass@1/5/10) on Instruments.}
\label{tab:ablation}
\end{table}

\subsubsection{\textbf{\textit{Overall Performance.}}}
We provide a detailed report in Table~\ref{tab:main_result} on the sequence recommendation performance of our method across three datasets. On the direct sequence recommendation task, our method achieves consistent improvements across all three datasets. On Instruments, Recall@10 of $\methodname_{Align}$ = 0.1207 surpasses EAGER-LLM = 0.1171 and LC-Rec = 0.1115; NDCG@10 likewise leads (0.0931 vs. 0.0890/0.0853). On Sports, NDCG@10 of $\methodname_{Align}$ = 0.0307 exceeds EAGER-LLM = 0.0293 and LC-Rec = 0.0247; on Beauty, Recall@5 of $\methodname_{Align}$ = 0.0567 also outperforms EAGER-LLM = 0.0534. Overall, Collaborative-Semantic Alignment substantially elevates retrieval quality and ranking effectiveness in the direct mode; moreover, because the sequential reasoning recommendation task and the direct task share a common semantic representation and curriculum signals, the two inference modes are mutually reinforcing. 

With reinforcement learning enabled ($\methodname_{RL}$), performance on the direct sequence recommendation task remains on par with, and in cases slightly superior to, the strongest baseline, while conferring policy-level stability and better generalization. For Instruments, Recall@10 of $\methodname_{RL}$ = 0.1139 is in the same range as EAGER-LLM = 0.1171, and NDCG@10 is similarly close (0.0884 vs. 0.0890). More importantly, on Beauty, despite small trade-offs at individual cutoffs, $\methodname_{RL}$ achieves a higher Recall@avg (arithmetic mean over Recall@{1,5,10}), 0.0498 > 0.0497 for EAGER-LLM, indicating that RL further consolidates overall recall without compromising direct-mode accuracy and provides a more controllable and robust policy foundation for subsequent reasoning-style generation.

On the sequential reasoning recommendation task, the effectiveness and robustness of SRPO are reflected in consistent gains in group success rates: on Instruments, Pass@1/5/10 improves from 0.0495/0.0705/0.0829 to 0.0650/0.0765/0.0845; on Beauty, Pass@1 rises from 0.0079 to 0.0137, and Pass@5 from 0.0270 to 0.0296, while Pass@10 shows a slight drop (0.0446 → 0.0403) at larger k, reflecting a stable exploration–exploitation trade-off. Coupled with our residual-sensitive verifiable reward and bonus-calibrated group advantage (combining dense–stable signals with rare–exact signals), SRPO mitigates variance collapse under sparse feedback and delivers robust improvements across a range of k thresholds.

\subsection{Ablation Study}

As show in~\ref{tab:ablation}, we conduct a staged ablation under a unified setup, evaluating both the \textit{Direct Sequence Recommendation} metric and the \textit{Sequential Reasoning Recommendation} metric. With only \textit{Collaborative-Semantic Alignment}, the Direct Sequence Recommendation metric reaches 0.0865. Adding \textit{Reasoning Curriculum Activation} further lifts it to 0.0949 (+0.0084, $\approx$\,+9.7\%) and, for the first time, yields a Sequential Reasoning Recommendation score of 0.0656. These results indicate that curriculum-guided reasoning supervision not only unlocks the model’s reasoning-mode capability but also provides a transferable representational boost to direct generation, suggesting that the combined alignment and curriculum supervision jointly strengthens collaborative semantics and causal reasoning representations.

Building on the curriculum-enhanced supervised fine-tuning model, we then dissect the reinforcement learning phase. \textit{Group Relative Policy Optimization} leads to a modest decrease in the Direct Sequence Recommendation metric from 0.0949 to 0.0921, while improving the Sequential Reasoning Recommendation metric from 0.0656 to 0.0741. The limited drop on the direct side is largely attributable to variance-guided dynamic sampling, which filters rollout groups with zero variance to avoid ineffective updates; while this improves stability, it also reduces the frequency of effective updates and thus dampens direct-side gains. Introducing the \textit{residual-sensitive verifiable reward} yields 0.0891 (Direct Sequence Recommendation) and 0.0724 (Sequential Reasoning Recommendation), a slight decline relative to Group Relative Policy Optimization—consistent with a prefix-aware concave dense reward that stabilizes coarse-grained correctness but exerts weaker pressure on exact hits. Augmenting with \textit{Bonus\mbox{-}Calibrated Group Advantage Estimation} recovers direct performance to 0.0899 and pushes reasoning to 0.0753, marginally surpassing Group Relative Policy Optimization (+0.0012 in the Sequential Reasoning Recommendation metric). Given current limits on the scale of pretraining and alignment data, we view the observed gains of Bonus-Calibrated Group Advantage Estimation as conservative; with larger datasets and stronger collaborative-semantic alignment, we expect its improvements—especially on reasoning-mode performance—to substantially outpace Group Relative Policy Optimization.
\begin{figure}[t]
  \centering
  \includegraphics[width=0.85\linewidth]{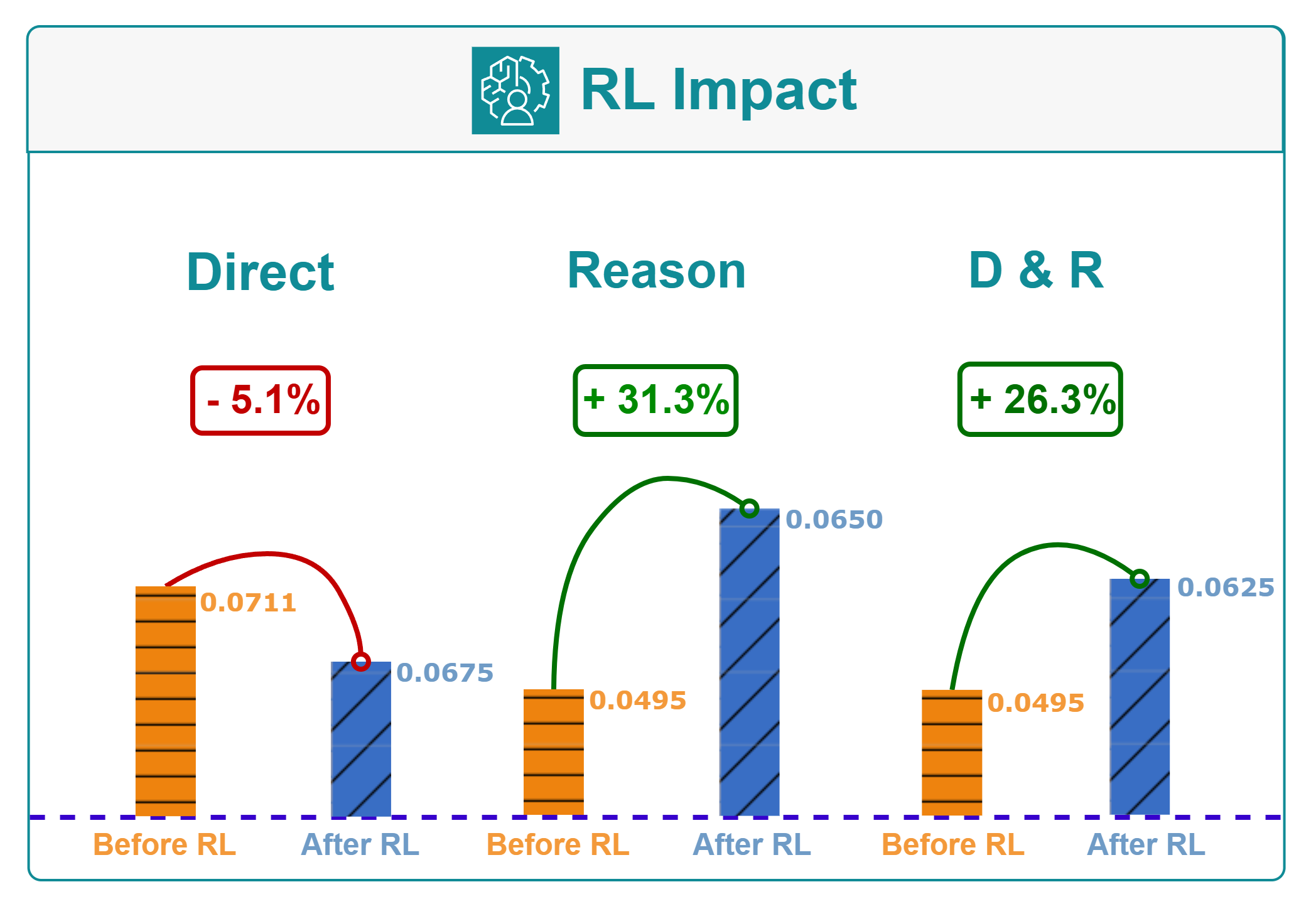}
  \caption{Impact of reinforcement learning (SRPO) across recommendation tasks on Instruments (Pass@1).}
  \label{fig:RL_ablation}
\end{figure}

\begin{table}[]
\centering
\resizebox{\columnwidth}{!}{%
\begin{tabular}{ccccc}
\hline
Datast Sourse & Direct & Reason & mixed D\&R & D -\textgreater R \\ \hline
\multicolumn{5}{c}{\cellcolor[HTML]{EFEFEF}Direct Sequence Recommendation} \\ \hline
\begin{tabular}[c]{@{}c@{}}Recall/NDCG\\ Avg.\end{tabular} & 0.0865 & 0.0789 & 0.0949 & 0.0718 \\ \hline
\multicolumn{5}{c}{\cellcolor[HTML]{EFEFEF}Sequential Reasoning Recommendation} \\ \hline
\begin{tabular}[c]{@{}c@{}}Pass@K\\ Avg.\end{tabular} & - & 0.0666 & 0.0677 & 0.0698 \\ \hline
\end{tabular}%
}
\caption{Effects of \emph{Collaborative\mbox{-}Semantic Alignment} and \emph{Reasoning Activation} data composition and training schedule on Direct Sequence Recommendation and Sequential Reasoning Recommendation.}
\label{tab:SFT_ablation}
\end{table}
\subsection{Scaling Law of Reasoning Activation}
As shown in Figure~\ref{fig:scaling_law}, after Collaborative–Semantic Alignment (without any curriculum), scaling \textit{Reasoning Activation}—measured by the amount of synthetic Chain-of-Thought (CoT) data and the corresponding training FLOPs—yields a stable, monotonic improvement in the average Pass@K metric (Pass@Avg). A linear-in-\(\log\) FLOPs fit achieves \(R^2=0.9549\), while a quadratic-in-\(\log\) FLOPs fit further improves to \(R^2=0.9734\). Across the explored range of \(10^{16}\)–\(10^{19}\) FLOPs, the curve rises almost linearly with a steady slope and \textbf{shows no sign of saturation or plateau}. The quadratic term introduces only mild curvature, yet the marginal gain remains positive throughout. These observations indicate that the current method has \textbf{not} reached its performance ceiling—the primary limitation is compute rather than methodological capacity. Consequently, under the same training recipe and evaluation protocol, increasing Reasoning Activation (i.e., more synthetic CoT data and training FLOPs) is expected to keep improving Pass@K.


\subsection{Further Analysis}
Table~\ref{tab:SFT_ablation} compares four supervised fine\mbox{-}tuning settings: {Direct} (training only {Collaborative\mbox{-}Semantic Alignment}); {Reason} (training only {Sequential Reasoning Recommendation} tasks without {Direct Sequence Recommendation} tasks); {mixed D\&R} (Curriculum training of alignment and reasoning); and {$D \rightarrow R$} (first train {Direct Sequence Recommendation} tasks, then {Sequential Reasoning Recommendation} tasks). Joint training achieves the best direct performance with Recall/NDCG of 0.0949 (a $\,+9.7\%$ improvement over the Direct baseline at $0.0865$), while also slightly exceeding the Reason\mbox{-}only setup on the reasoning metric (Pass@K average $0.0677$ vs.\ $0.0666$), indicating positive transfer and shared representations between alignment and reasoning objectives. The sequential schedule $D \rightarrow R$ yields the highest reasoning score (0.0698) but reduces the direct metric to $0.0718$, revealing forgetting when switching objectives without replay or regularization. Training only reasoning notably lowers direct performance ($0.0789$), underscoring the necessity of collaborative\mbox{-}semantic alignment for maintaining retrieval and ranking capacity. Overall, mixed training provides a stronger balance between efficiency/accuracy (direct) and explainable reasoning; if maximizing reasoning is the priority, $D \rightarrow R$ can be employed, ideally with interleaved sampling, experience replay, or KL regularization to mitigate negative transfer on the direct objective.

Figure~\ref{fig:RL_ablation} evaluates Sparse-Regularized Group Policy Optimization on the Instruments dataset across three training regimes. Applying reinforcement learning only to Direct Sequence Recommendation reduces Pass@1 from 0.0711 to 0.0675 (-5.1\%), suggesting limited benefit for direct decoding under this signal. In contrast, training only on Sequential Reasoning Recommendation lifts Pass@1 from 0.0495 to 0.0650 (+31.3\%), indicating that verifiable rewards primarily strengthen the reasoning pathway. Joint training on Direct\&Reason improves reasoning to 0.0625 (+26.3\%) but underperforms the reasoning-only schedule, consistent with cross-objective interference when capacity is shared.

\section{Conclusion}

We presented a unified LLM-based generative reasoning recommendation framework that integrates Collaborative-Semantic Alignment, Reasoning Curriculum Activation, and Sparse-Regularized Group Policy Optimization (SRPO). By bridging the semantic gap between language modeling and collaborative interaction data, our approach enables both efficient direct recommendation and interpretable reasoning-based generation under verifiable reinforcement signals. The proposed framework not only advances verifiable-RL-driven recommendation but also provides a practical path toward transparent, causally grounded, and scalable LLM recommender systems.



\bibliographystyle{ACM-Reference-Format}
\bibliography{sample-base}

\appendix

\section{Limitation}
While our framework demonstrates consistent improvements across benchmarks, the current study is bounded by computational and resource constraints. The scaling trajectory of reasoning performance indicates that the model remains within the non-saturated regime of the scaling law, suggesting further potential with larger compute budgets. Due to these limitations, we have not yet explored training with larger backbones or extended reasoning curricula, which may yield additional gains in semantic alignment and reinforcement learning stability. Future work will investigate these directions to fully exploit the scalability and generalization potential of our approach.

\section{Experimental Supplement}
\subsection{Baselines.}
\label{baseline}
We compare the following four categories of methods:

\noindent (1) \emph{Traditional sequential methods}
\begin{itemize}[leftmargin=*]
    \item \textbf{GRU4REC} \cite{hidasi2015gru4rec}: An RNN-based model that employs GRUs to capture sequential dependencies in user–item interactions.
    \item \textbf{Caser} \cite{tang2018caser}: A CNN-based model that learns sequential patterns through horizontal and vertical convolutional operations.
    \item \textbf{HGN} \cite{ma2019hgn}:  A hierarchical gating network that models both long-term and short-term user preferences.
\end{itemize}

\noindent (2) For \emph{transformer-based methods}, we have:
\begin{itemize}[leftmargin=*]
    \item \textbf{S\textasciicircum{}3-Rec} \cite{zhou2020s3}: Enhances sequential recommendation by pre-training a bidirectional Transformer to maximize mutual information.
    \item \textbf{BERT4Rec} \cite{sun2019bert4rec}:  Utilizes a bidirectional Transformer architecture to capture contextual item relationships beyond unidirectional models.
    \item \textbf{FDSA} \cite{zhang2019feature}: Applies self-attention to model transitions among feature-level item representations.
\end{itemize}

\noindent (3) For \emph{generative methods}, we have:
\begin{itemize}[leftmargin=*]
    \item \textbf{P5-CID} \cite{hua2023p5_cid}: A T5-based generative recommender that constructs collaborative ID representations to enhance cross-domain generalization.
    \item \textbf{TIGER} \cite{rajput2024tiger}: Leverages a T5 backbone to autoregressively generate target item identifiers through sequential decoding.
\end{itemize}

\noindent (4) For \emph{LLM-Based methods}, we have:
\begin{itemize}[leftmargin=*]
    \item \textbf{LC-Rec} \cite{zheng2024lcrec}: Incorporates vector quantization to learn semantic identifiers and employs LLaMA for autoregressive item prediction.
    \item \textbf{EAGER-LLM} \cite{hong2025eager}: A decoder-only llm-based generative recommendation framework that integrates endogenous and exogenous behavioral and semantic information in a non-intrusive manner. 
\end{itemize}

\section{Case Study}
\label{casestudy}

We show a case study that our model successfully performs reasoning first, then recommends the item below.

\begin{tcolorbox}[
    colback=gray!5!white,      
    colframe=gray!60!black,    
    title=Case Study - Instruments, 
    fonttitle=\bfseries,       
    breakable,
]

\textbf{System}: 

You are an AI recommendation model. You task is to analyze the user's purchase history to recommend what they would likely purchase next.

Your output must be structured into following format:
\newline
\texttt{<think>}
\begin{itemize}
    \setlength\itemsep{0em} 
    \item \textbf{Step 1: Extraction of Behavioral Evidence:} Examine the items in the purchase history, including their metadata, to identify patterns, trends, or any other insights.
    \item \textbf{Step 2: Modeling of Latent Preferences:} Based on the analysis, infer and summarize the user's preferences.
    \item \textbf{Step 3: Inference of User Intent and Goal Formulation:} Deduce the user's potential purchase intent or next likely item by connecting the history and preferences.
\end{itemize}
\texttt{</think>}
\newline
Then rewrite the user purchase history and recommend the next item the user may buy based on your analysis. Enclose the item in \texttt{<answer> </answer>} xml tags.

\textbf{User}: The user has interacted with items \{user purchase history\} in chronological order. Can you predict the next possible item that the user may expect?

\textbf{Assistant}:

\texttt{<think>} \newline
\textbf{Step 1: Extraction of Behavioral Evidence:}

The user has purchased four items related to music and audio production. Two items are microphones and audio equipment (Blue Snowball USB Microphone and JOYO JA-03 Tube Drive Sound Effect Mini Guitar Pocket Amplifier), one is a guitar part (Musiclily Plastic 3 Way Pickup Selector Toggle Switch), and one is a guitar amplifier (Fender Mustang I V2 20-Watt 1x8-Inch Combo Electric Guitar Amplifier). The user seems to be interested in music production, guitar playing, and audio equipment.

\textbf{Step 2: Modeling of Latent Preferences:}

Based on the purchase history, the user appears to have the following preferences:
\begin{enumerate}
    \setlength\itemsep{0em} 
    \item Interest in music production and audio equipment.
    \item Preference for high-quality, professional-grade products.
    \item Focus on guitar playing and guitar-related accessories.
    \item Desire for compact, portable, and easy-to-use products.
\end{enumerate}

\textbf{Step 3: Inference of User Intent and Goal Formulation:}

The user's purchase history suggests a pattern of buying products that enhance their music production and guitar playing experiences. They seem to be looking for products that are easy to use, compact, and of high quality. The user may be interested in upgrading or modifying their existing equipment to improve their sound or playing experience. \newline
\texttt{</think>}

Let's rewrite user's purchase history: \{user purchase history\}.

Based on the above analysis, the recommended product is \texttt{<answer>}<|a\_114|><|b\_202|><|c\_184|><|d\_188|> \texttt{</answer>}
\end{tcolorbox}

\begin{tcolorbox}[
    colback=gray!5!white,      
    colframe=gray!60!black,    
    title=Case Study - Beauty, 
    fonttitle=\bfseries,       
    breakable,
]

\textbf{System}: 

You are an AI recommendation model. You task is to analyze the user's purchase history to recommend what they would likely purchase next.

Your output must be structured into following format:
\newline
\texttt{<think>}
\begin{itemize}
    \setlength\itemsep{0em} 
    \item \textbf{Step 1: Extraction of Behavioral Evidence:} Examine the items in the purchase history, including their metadata, to identify patterns, trends, or any other insights.
    \item \textbf{Step 2: Modeling of Latent Preferences:} Based on the analysis, infer and summarize the user's preferences.
    \item \textbf{Step 3: Inference of User Intent and Goal Formulation:} Deduce the user's potential purchase intent or next likely item by connecting the history and preferences.
\end{itemize}
\texttt{</think>}
\newline
Then rewrite the user purchase history and recommend the next item the user may buy based on your analysis. Enclose the item in \texttt{<answer> </answer>} xml tags.

\textbf{User}: The user has interacted with items \{user purchase history\} in chronological order. Can you predict the next possible item that the user may expect?

\textbf{Assistant}:

\texttt{<think>} \newline
\textbf{Step 1: Extraction of Behavioral Evidence:}

\begin{itemize}
    \setlength\itemsep{0em}
    \item Product types: Hydrating serum (hyaluronic acid), facial oils (rosehip, argan, emu), eye gel, anti-aging moisturizer.
    \item Key concerns: Dryness, fine lines, wrinkles, dark circles, puffiness, uneven tone/age spots; emphasis on anti-aging and deep hydration.
    \item Ingredient preferences: Natural/organic, cold-pressed, unrefined oils; botanicals (aloe, cucumber, green tea); hyaluronic acid; vitamin E; gentle, non-greasy formulas.
    \item Brand/assurance: Repeated preference for InstaNatural and products with strong money-back guarantees.
    \item Routine coverage: Has serum, moisturizer, eye treatment, and multiple oils; missing a dedicated gentle cleanser and daily SPF.
\end{itemize}

\textbf{Step 2: Modeling of Latent Preferences:}

\begin{itemize}
\setlength\itemsep{0em}
\item Natural, clean, and gentle formulations suitable for sensitive/dry skin.
\item Multi-benefit, anti-aging products that hydrate, brighten, and improve texture.
\item Trusted brands with satisfaction guarantees (notably InstaNatural).
\item Lightweight, fast-absorbing, non-greasy textures.
\item Value in larger sizes or long-lasting products.
\end{itemize}

\textbf{Step 3: Inference of User Intent and Goal Formulation:}

The user is building a comprehensive anti-aging, hydrating routine. With serum, moisturizer, eye gel, and oils in place, the next logical addition is a gentle, natural cleanser to prep skin without stripping moisture, ideally with soothing botanicals and hyaluronic acid to align with their preferences and brand loyalty. \newline
\texttt{</think>}

Let's rewrite user's purchase history: \{user purchase history\}.

Based on the above analysis, the recommended product is \texttt{<answer>}<|a\_125|><|b\_109|><|c\_135|><|d\_125|> \texttt{</answer>}
\end{tcolorbox}

\begin{tcolorbox}[
    colback=gray!5!white,      
    colframe=gray!60!black,    
    title=Case Study - Sports, 
    fonttitle=\bfseries,       
    breakable,
]

\textbf{System}: 

You are an AI recommendation model. You task is to analyze the user's purchase history to recommend what they would likely purchase next.

Your output must be structured into following format:
\newline
\texttt{<think>}
\begin{itemize}
    \setlength\itemsep{0em} 
    \item \textbf{Step 1: Extraction of Behavioral Evidence:} Examine the items in the purchase history, including their metadata, to identify patterns, trends, or any other insights.
    \item \textbf{Step 2: Modeling of Latent Preferences:} Based on the analysis, infer and summarize the user's preferences.
    \item \textbf{Step 3: Inference of User Intent and Goal Formulation:} Deduce the user's potential purchase intent or next likely item by connecting the history and preferences.
\end{itemize}
\texttt{</think>}
\newline
Then rewrite the user purchase history and recommend the next item the user may buy based on your analysis. Enclose the item in \texttt{<answer> </answer>} xml tags.

\textbf{User}: The user has interacted with items \{user purchase history\} in chronological order. Can you predict the next possible item that the user may expect?

\textbf{Assistant}:

\texttt{<think>} \newline
\textbf{Step 1: Extraction of Behavioral Evidence:}

\begin{itemize}
    \setlength\itemsep{0em} 
\item The user consistently buys compact, affordable home fitness equipment suited for small spaces (doorway pull-up bar, adjustable dumbbell bars, mini hoop, fitness ball, roller slide).
\item Emphasis on bodyweight and functional training tools (pull-ups, dips, push-ups, core work, light strength).
\item Preference for items that are easy to set up, store, and travel with (door-mounted, folding, portable, no permanent installation).
\item Interest in guided use and instruction (exercise charts, DVDs, printed exercise guide on ball).
\item Focus on upper body and core development, with versatility and progression (adjustable dumbbell bars suggest scalable resistance; multiple grip options; varied workout modes).
\end{itemize}

\textbf{Step 2: Modeling of Latent Preferences:}

Based on the purchase history, the user appears to have the following preferences:
\begin{itemize}
    \setlength\itemsep{0em} 
\item Compact, space-saving equipment that fits in a home or office environment.
\item Bodyweight and functional training with options to scale difficulty.
\item Easy setup, portability, and minimal installation.
\item Value-oriented, durable materials (steel, sturdy construction).
\item Tools that enhance core strength, upper body, and overall fitness with included guidance.
\end{itemize}

\textbf{Step 3: Inference of User Intent and Goal Formulation:}

\begin{itemize}
    \setlength\itemsep{0em} 
\item The user likely seeks to expand their bodyweight training with more versatile, scalable resistance and better progression for push-ups, dips, and core movements.
\item A logical next step is a tool that complements existing equipment, increases challenge, and remains compact and portable.
\item An adjustable push-up/dip system with rotating handles aligns with their goals, enabling varied grips, deeper range of motion, and scalable resistance.
\end{itemize}
\texttt{</think>}

Let's rewrite user's purchase history: \{user purchase history\}.

Based on the above analysis, the recommended product is \texttt{<answer>}<|a\_210|><|b\_26|><|c\_249|><|d\_204|> \texttt{</answer>}
\end{tcolorbox}

\end{document}